\documentclass{JHEP}

\usepackage{epsfig,bbm,bm,amssymb}

\def\nc{N_{\rm c}}
\def\Dp{\Delta_p}
\def\Dk{\Delta_k}
\def\Dq{\Delta_{k{+}p}}
\def\hfp{\frac{p}{2}}
\def\hfq{\frac{k{+}p}{2}}
\def\hfk{\frac{k}{2}}

\def\eps{\varepsilon}

\def\Tr{\,{\rm Tr}\:}

\def\hc{{\rm h.c.}}

\def\df{d_{\rm f}}

\def\hf{{1\over 2}}

\def\st{\begin{equation}}
\def\stp{\end{equation}}
\def\bg{\begin{eqnarray}}
\def\nd{\end{eqnarray}}
\def\nn{\nonumber}

\def\Dsl{\hbox{/\kern-.6000em\rm D}}
\def\dsl{\hbox{/\kern-.5600em$\partial$}}
\def\pxpsl{\hbox{/\kern-.5600em$p$}}
\def\ssl{\hbox{/\kern-.5600em$s$}}
\def\epssl{\hbox{/\kern-.5600em$\epsilon$}}
\def\delsl{\hbox{/\kern-.7000em$\nabla$}}
\def\lxpsl{\hbox{/\kern-.5600em$l$}}
\def\kxpsl{\hbox{/\kern-.5600em$k$}}
\def\qxpsl{\hbox{/\kern-.5600em$q$}}

\def\L{{\cal L}}

\def\hf{{1\over 2}}
\def\g1{$U_Y(1)$}
\def\g2{$SU_L(2)$}
\def\g3{$SU_c(3)$}
\def\g31{$SU_c(3)\times U_{em}(1)$}
\def\g21{$SU_L(2)\times U_Y(1)$}
\def\g321{$SU_c(3)\times SU_L(2)\times U_Y(1)$}

\def\nott#1{\setbox0=\hbox{$#1$}                
   \dimen0=\wd0                                 
   \setbox1=\hbox{/} \dimen1=\wd1               
   \ifdim\dimen0>\dimen1                        
      \rlap{\hbox to \dimen0{\hfil/\hfil}}      
      #1                                        
   \else                                        
      \rlap{\hbox to \dimen1{\hfil$#1$\hfil}}   
      /                                         
   \fi}                                         %

\input epsf

\author{Joshua W.\ Elliott and Guy D.\ Moore\\
 McGill University,\\
 3600 University St.,\\
 Montr\'{e}al, QC H3A 2T8, Canada\\
 email: elliottj@physics.mcgill.ca, guymoore@physics.mcgill.ca
	}

\title{Three Dimensional N=2 Supersymmetry on the Lattice}
\date {\today}

\abstract%
    {%
We show how 3-dimensional, $N{=}2$ supersymmetric theories, including
super QCD with matter fields, can be put on the lattice with existing
techniques, in a way which will recover supersymmetry in the small
lattice spacing limit.  Residual supersymmetry breaking effects are
suppressed in
the small lattice spacing limit by at least one power of the lattice
spacing $a$.
    }%

\preprint{...}

\begin{document}

\section{Introduction}

Supersymmetry is a beautiful and phenomenologically interesting
candidate for phys{-}ics beyond the standard model.  Our
understanding of quantum field theory has been improved by our ability
to make exact statements within supersymmetric theories, such as
non-renormalization theorems and dualities. 

There is a large class of nonperturbative results or conjectured results
for supersymmetric theories, particularly supersymmetric Yang-Mills
theories, both in four dimensions
\cite{Intriligator:1995au,Seiberg:1994aj,Poppitz:1998vd} and in three
dimensions \cite{Intriligator:1996ex,Strassler,deBoer,Hanany:1996ie}. 
For instance, in 4 dimensional $N{=}2$ 
super Yang-Mills (SYM) theory, Seiberg and
Witten's work \cite{Seiberg:1994aj} essentially solves the theory.  In
three dimensions, a number of exquisite results have been obtained
involving mirror symmetry in
$N{=}4$ SYM theory \cite{Intriligator:1996ex} 
and also in 
$N{=}2$ theories \cite{Strassler,deBoer}. In \cite{Hanany:1996ie}
these symmetries were motivated in the context of string theory. 
There is also a beautiful conjecture by Witten \cite{Witten:N=1} which
makes $N{=}1$ SYM theory in 3 dimensions an excellent testbed
for spontaneous (dynamical) supersymmetry breaking.

It would be very helpful if we were able to test some of the techniques
for studying supersymmetric theories, for instance by ``solving'' the
theories involved in a non-perturbative way.  The lattice is the best
candidate method for solving such theories.  Unfortunately, the
lattice regulator almost inevitably breaks the supersymmetry. The 
lattice is, after all, a regularization scheme designed to preserve 
exact gauge symmetry at the expense of manifest Poincar\'e invariance. 
Since supersymmetry is a space-time symmetry, i.e. an extension of 
the Poincar\'e group, it is of no surprise that it is broken on the 
lattice. 

It is certainly possible to use the lattice to study a theory which
possesses a symmetry broken by the lattice action.  After all, the main
application of the lattice technique is the study of QCD, which has O(4)
Euclidean invariance, broken to (hyper)cubic symmetry on the lattice.
In this case, the symmetry is recovered automatically, up to corrections
suppressed by the lattice spacing, because the low-energy phenomena are
described by an effective theory which possesses full O(4) symmetry
accidentally.  That is, the low dimension operators which respect
(hyper)cubic symmetry also respect O(4) symmetry, so that symmetry is
recovered in the infrared.  Unfortunately, for supersymmetric theories
containing scalars, the scalar mass term is low dimensional and breaks
the supersymmetry.  Therefore very fine tuning of the lattice action is
generally required to recover supersymmetry in the infrared.

A great
deal of work has gone into looking for a way to preserve supersymmetry,
at least in the small lattice spacing limit
\cite{Golterman,Catterall,kaplan,other_WZ2D,Fujikawa}. 
In 4 dimensions, it appears that one should be able to implement $N{=}1$
SYM theory without matter \cite{Kaplan4D,others,4DSUSY},
that is, a theory containing only gauge fields and an adjoint
representation, Weyl fermion (the gluino).  
This is because this case does not involve a scalar field, so
supersymmetry is an accidental symmetry once enough chiral
symmetry is present to prevent a mass for the gluino.  This is now
possible with recent advances in 4-D chiral fermions
\cite{Kaplan_domainwall,Narayanan,Luscher,others2}. 
For cases involving scalar fields, most attempts involve building in
enough supersymmetry into the lattice construction to ensure that scalar
mass renormalization does not occur.  This is true, for instance, of
proposed implementations of the Wess-Zumino model in 2 dimensions
\cite{Golterman,Catterall,Giedt:2005ae}.  Various
2 and 3 dimensional, highly supersymmetric pure Yang-Mills theories can
be implemented by a technique developed by Kaplan, Katz, and Unsal
\cite{kaplan,kaplan2} (see however \cite{Giedt}). A technique for 
putting $N{=}2$ SYM theory on the lattice using ``twisted''
supersymmetries  with K\"ahler-Dirac fermions has been developed by Catterall 
\cite{Catterall:2004np} and recently extended to $N{=}4$ supersymmetry
in 4 dimensions
\cite{Catterall:2005fd}. Also D'Adda and collaborators \cite{D'Adda:2005zk} 
have recently 
completed the construction of a 2-d lattice which retains {\em all} 
the supersymmetries of the target SYM theory exactly on the lattice 
using twisted supersymmetry and ``fermionic'' links where supercharges 
live in direct analogy to gauge fields living on bosonic links. This 
would seem to be well motivated by analogy to the continuous superspace 
formulation of supersymmetry in which the generators of the 
supersymmetry are regarded as translation operators along new 
Grassman valued coordinates. 

These techniques often come with certain limitations such as confinement 
to even dimensions, to very highly supersymmetric theories or to 
specific numbers of fermionic fields. They are 
frequently made difficult by highly non-trivial 
lattice actions that can contain 
a wide range of complications such as non-commutativity, the failure of 
reflection positivity, unphysical moduli, or sign problems with complex 
fermion determinants. Most of
these issues can or have been overcome to varying extents,
yet they still remain
to complicate the treatment.

We argue here that, for all the 3-dimensional theories of interest with
$N{=}2$ (4 real supercharges) or more, one can proceed by actually doing
the fine tuning on the most conventional lattice action available, the
Wilson action with Wilson fermions.  The reason this approach is
available is because these theories are super-renormalizable.  That
means that loop corrections involving the ultraviolet converge in powers
of the lattice spacing, so the tuning only requires a lattice
perturbation theory calculation to a finite loop order, which turns out
to be 2 loops.  Further, we explicitly do this calculation for the case
of SYM theory plus fundamental matter.
This holds out the possibility of testing some very
interesting claims \cite{Strassler,deBoer} 
of exact results in $N{=}2$ SYM theory with matter. Some 
similarly spirited work in the context of 2-D $N{=}2$ pure SYM has recently 
appeared in \cite{Suzuki:2005dx}.

In Section \ref{sec:idea}, we will explain the general idea which makes it
possible to preserve supersymmetry, in the small lattice spacing limit,
in 3-dimensional, $N{=}2$ supersymmetric theories.  In Section
\ref{sec:WessZumino} we present in detail the implementation of the
Wess-Zumino model, that is, a theory of fermions and scalars.  In
Section \ref{sec:gauge} we present the implementation of gauge theories.
Finally, some concluding remarks appear in Section \ref{sec:conclusion}.

\section{Super-renormalizability and SUSY breaking}
\label{sec:idea}

No one is interested in the results of a lattice calculation {\it per
se}.  After all, a lattice ``field theory'' is actually just a
statistical model, not a true field theory.  The reason that the lattice
technique can teach us something about field theory, is that in the {\em
infrared}, the correct effective description of a lattice theory is as a
quantum field theory.  What one must do is to ensure that the infrared
behavior of the lattice theory coincides with the continuum quantum
field theory of interest, or at least that it does so in the small
lattice spacing limit, so that the behavior of the field theory can be
probed by making a zero lattice spacing extrapolation.  

It is easy to write down a lattice gauge theory which, at tree level,
will look in the infrared like the theory of interest%
\footnote
    {%
    In four dimensions this statement is true of vectorlike theories,
    but serious complications arise if one wants a chiral theory, that
    is, a theory with two component spinors in a representation which is
    not real or pseudoreal and which are not balanced by an equal number
    of spinors of opposite handedness in the same representation.  To
    date, it is understood how, in principle, to put certain Abelian
    chiral theories on the lattice; it it not clear, even in principle,
    how to put more general theories on the lattice while still 
    retaining manifest gauge invariance \cite{Luscher_chiral}.
    Fortunately, for the three dimensional theories of interest to us,
    this is not relevant, because chiral fermions do not exist in odd
    dimensions.
    }.
The problem is that, in the UV (at the lattice spacing scale), the
lattice theory typically does not have the full symmetries of the theory
we are interested in.  Generally it is possible to formulate lattice
theories so that they have exact gauge and (hyper)cubic symmetries.
However, under supersymmetry, the variation of a fermionic field can
involve the derivative of a bosonic field; and since derivatives become
finite differences on the lattice, 
supersymmetry will generically be badly
broken at the lattice spacing scale.  Furthermore, even if we construct
the lattice theory to satisfy supersymmetric relations in the infrared,
radiative effects involving UV (SUSY breaking) modes will typically
communicate those effects to the infrared modes of interest.

The IR effective theory is not the tree level theory.  Rather, it is the
theory one obtains, by writing down the most general continuum quantum
field theory consistent with the field content and symmetries of the
lattice, and performing a matching calculation between the lattice
theory and that continuum effective theory, to determine what the actual
parameters of the IR effective theory are.  For instance, if we made a
tree level lattice implementation of the Wess-Zumino model (which in 3
dimensions exhibits $N{=}2$ supersymmetry, that is, it has 4 real
supersymmetry generators),
\begin{equation}
\L_{\rm bare} = \partial_\mu \Phi^* \partial^\mu \Phi 
        + \psi^\dag \nott{\partial} \psi
        + \Big( \lambda \Phi \psi^{\!\top} e \psi + \hc \Big) 
        + \lambda^2 \Big( \Phi^\dagger \Phi \Big)^2 \, ,
\label{eq:Lbare}
\end{equation}
with $\Phi$ a complex scalar and $\psi$ a two component spinor, then we
would generically recover an infrared theory where all terms
permissible with this field content were present;
\begin{eqnarray}
\L_{\rm IR}  & = &  Z_\phi \partial_\mu \Phi^* \partial^\mu \Phi 
        + Z_\psi \psi^\dag \nott{\partial} \psi
        + m_\phi^2 \Phi^* \Phi
        + m_\psi \psi^\dag \psi 
\nonumber \\ && {}
        + \Big( \lambda_y \Phi \psi^{\!\top} e \psi + \hc \Big) 
        + \lambda_s^2 \Big( \Phi^* \Phi \Big)^2
        + \mbox{ (High Dim.)}\, .
\label{eq:eff_th}
\end{eqnarray}
Here $Z_\phi$ and $Z_\psi$ represent the difference in field
normalization between the lattice and continuum fields; they can be
removed by a field rescaling, but we must keep them in mind when we
compare lattice correlation functions with their continuum counterparts.
The point is that the IR behavior typically involves radiatively
generated terms which do not respect the intended supersymmetry.  In
particular one does not expect $m_\psi^2 = m_\phi^2$. 

In 4 dimensions this problem is severe.  The SUSY violating, radiatively
induced terms appear at all orders in perturbation theory, with
coefficients, at high order, which are only suppressed with respect to
the lower order coefficients by powers of a dimensionless coupling.
Further, additive scalar mass renormalizations are divergently large
at {\em every loop order}.  That is, in 4 dimensions, the contributions
to the mass squared parameter are of order
\begin{equation}
\delta m^2 \; \mbox{at} \qquad
\mbox{1 loop: } \lambda^2/a^2 \, ; \quad
\mbox{2 loops: } \lambda^4/a^2 \, ; \quad
\mbox{3 loops: } \lambda^6/a^2 \, ; \; \ldots \, ,
\end{equation}
where $a$ is the lattice spacing.
Every such coefficient is problematic; a
severe nonperturbative tuning is needed to remove them.  It is not at
all clear how to perform such a tuning; generally we can only perform
nonperturbative tunings in lattice gauge theories if we have one exact
conservation law or Ward identity per tuning required.

The beauty of 3-D is that the desired theory is generally
super-renormalizable.  Consequently, the UV is very weakly coupled;
specifically, as the lattice spacing is taken to zero, the coupling
at the scale of the lattice spacing falls linearly with lattice spacing
$a$. This means that, while the SUSY breaking nature of
the UV regulator radiatively induces SUSY breaking
effects in the IR, the matching calculation which
determines them converges very quickly.  At each loop order, we
determine the matching of parameters to one more power of the lattice
spacing $a$.  For instance, in the above model, if we compute the mass
squared for the scalar field, generated by UV physics, the contributions
at different orders in the loopwise expansion are again of order $\lambda^2$,
$\lambda^4$, $\lambda^6$, $\ldots$.  But $\lambda^2$ has mass dimension
1.  Since the matching calculation involves only UV physics, the only
scale which can balance the explicit powers of mass is the lattice
spacing scale.  Therefore, the terms in the loopwise expansion
are of order 
\begin{equation}
\delta m^2 \; \mbox{at} \qquad
\mbox{1 loop: } \lambda^2/a \, ; \quad
\mbox{2 loops: } \lambda^4 \, ; \quad
\mbox{3 loops: } a \lambda^6 \, ; \; \ldots \, .
\end{equation}
The one and two loop contributions are significant and must be removed
by an appropriate counterterm.  However, three and higher loop effects
vanish in the $a \rightarrow 0$ limit, and so can be neglected.
For the scalar self-coupling $\lambda_s^2$, the one loop correction is
already $O(a \lambda^4)$, and so a tree level treatment is already
sufficient. 

We should stress here that the infrared cancellation which ensures that
only the scale $a$ can appear to balance the explicit mass dimensions in
the coupling, is a generic property of matching 
calculations in effective field theory and contains no statement about 
the IR physics of supersymmetry. Infrared divergences arise from large 
length scale (or low loop momentum) behavior. By construction 
the two theories for which 
any matching in an effective field theory formulation 
is being performed have the same IR behavior, and so any IR 
divergence will cancel in the difference. The statement 
is then simply that the IR extension 
of the lattice theory does describe the SUSY theory of interest 
provided that it contains the same degrees of freedom and that the 
coefficients of the terms in the Lagrangian can be matched with those 
of the theory of interest. In the 4-d case this matching does not seem 
possible within a perturbative framework; however, in a 
super-renormalizable theory it certainly is. 

We see that only a finite loop order is needed before all
remaining corrections are suppressed by powers of $a$.  It is therefore
feasible to perform the matching calculation to the requisite order
analytically, and to tune
the lattice theory based on the {\em purely analytic} result of this
perturbative matching calculation, to ensure that the IR effective
theory satisfies all relations implied by SUSY up to $a$ suppressed
corrections.

How hard is this tuning?  In the present paper we will be satisfied with
removing SUSY violating effects which do not vanish with the lattice
spacing; that is, we will leave $O(a)$ SUSY violating effects, but
prevent $O(a^0)$ SUSY violation.  In this case it is straightforward to
show that the scalar 
masses must be determined at two loops, and the fermionic masses must be
determined at one loop.  All other Lagrangian parameters may be taken at tree
level.  
This same technique has been applied to purely bosonic
3-dimensional Yang-Mills Higgs theory, in the context of understanding
the electroweak phase transition, in
\cite{FKRS2,Laine,Oa1,LaineRajantie,Oa2}.  Note that we will not attempt
to provide renormalized operators, and in particular we will not compute
the SUSY violating renormalization of the vacuum energy, which would
require a 4 loop effort.

To conduct the matching calculation, we need to compute some set of
correlation functions in both the lattice theory and the continuum,
supersymmetric theory, and equate the answers.  At the level of
interest, we need only do so for two correlation functions, which must
be sensitive to the scalar and fermionic masses.  The obvious candidates
are the respective two-point functions (self-energies).  Further, there
is no need to compute the supersymmetric values, since we already know
that they vanish identically.  Therefore, all that is required is to
compute the fermionic self-energy at zero momentum, at one loop, and the
scalar self-energy at zero momentum, at two loops, and to assign
counterterms to cancel these contributions.  Provided that we take
suitable combinations of loop contributions (fermionic and bosonic), we
are guaranteed that all loop integrations will be IR finite--precisely
because supersymmetry is satisfied in the infrared.  Therefore, the
matching calculation will consist of computing a handful of IR finite
linear combinations of Feynman graphs involving scalars, fermions, and
gauge fields.

In our calculations here, we will eliminate all corrections which are
unsuppressed by powers of the lattice spacing $a$; but corrections
suppressed by a single power of $a$ will still exist.
If we were very strong, it would be possible to eliminate $O(a)$ errors
as well, by conducting a 1 loop determination of couplings, a 2 loop
determination of fermionic masses, and a 3 loop determination of scalar
masses.  (Tree level improvement of the fermionic action would also be
required.)  Such a calculation has been carried out for three
dimensional, $O(N)$ symmetric
scalar field theories \cite{ArnoldMoore}; indeed, in this case many of
the $O(a^2)$ corrections have also been eliminated \cite{Tranberg}.
In principle it is possible to eliminate errors to any fixed order in
$a$, by using a suitably tree-level improved action and evaluating a
number of low dimension operators to suitable loop levels; as the number
of powers of $a$ accuracy desired increases, more and more operators
must be improved.  We will not attempt such a program here.

\section{Detailed treatment of the Wess-Zumino model}
\label{sec:WessZumino}

Consider a theory of scalars and fermions with $N{=}2$ supersymmetry in 3
dimensions.  The matter content is the same as for a 4 dimensional $N{=}1$
supersymmetric theory.  To make the extension to gauge theories simpler,
we consider a theory with a {\em global} SU($\nc$) symmetry, and three
sets of fields; those in the fundamental representation, $\Phi_f$,
$\psi_f$; those in the antifundamental representation, $\Phi_a$,
$\psi_a$; and those in the singlet representation, $\Phi_s$, $\psi_s$.

The most general superpotential, excluding mass terms, is
\st
W = \lambda_{ijk} \Phi_{f,i} \Phi_{a,j} \Phi_{s,k}
+ \frac{\xi_{ijk}}{6} \Phi_{s,i} \Phi_{s,j} \Phi_{s,k} \, ,
\stp
with $\xi_{ijk}$ totally symmetric in its indices.  We have left out
supersymmetric 
mass terms simply because including them will not lead to any change in
the mass counterterms we will need; any loopwise correction involving a
mass term will be $O(m^2)$; at worst this leads to $O(m^2 \lambda^2 a)$
effects, which are beyond the order under consideration.

The Lagrangian which follows from this superpotential is
\bg
{\cal L} & = &\bigg[ 
	(\partial_\mu \Phi_{f,i})^* (\partial^\mu \Phi_{f,i})
        + \psi^\dag_{f,i} \, \nott{\partial} \psi_{f,i}
+(f\to a,s) \bigg]
        + \lambda^*_{ijk} \lambda_{ilm} \Phi^*_{a,j} \!\cdot \Phi_{a,l}
        \Phi^*_{s,k} \Phi_{s,m}\nn \\ &&
        + \lambda^*_{jik} \lambda_{lim} \Phi^*_{f,j} \!\cdot \Phi_{f,l}
        \Phi^*_{s,k} \Phi_{s,m}
        + \lambda^*_{jki} \lambda_{lmi} \Phi^*_{f,j} \!\cdot \Phi^*_{a,k} 
        \Phi_{f,l} \!\cdot \Phi_{a,m}
\nn \\ &&
        + \left( \frac{\lambda_{jki} \xi^*_{lmi}}{2} \Phi_{f,j} \!\!\cdot
        \Phi_{a,k} \Phi^*_{s,l} \Phi^*_{s,m} \; + \hc \right)
        + \frac{\xi_{ijk} \xi^*_{ilm}}{4} \Phi_{s,j} \Phi_{s,k}
        \Phi^*_{s,l} \Phi^*_{s,m}
\nn \\ &&
        + \Bigg( \lambda_{ijk}\Big[ 
         \Phi_{s,k} \psi^\top_{f,i} \,e\, \psi_{a,j}
        +\Phi_{f,i} \psi^\top_{s,k} \,e\, \psi_{a,j}
        +\Phi_{a,j} \psi^\top_{s,k} \,e\, \psi_{f,i}
        \Big]
\nn \\ && \;\;\,
        + \lambda^*_{ijk} \Big[
         \Phi^*_{s,k} \psi^\dag_{f,i} \,e\, \psi^*_{a,j}
        +\Phi^*_{f,i} \psi^\dag_{s,k} \,e\, \psi^*_{a,j}
        +\Phi^*_{a,j} \psi^\dag_{s,k} \,e\, \psi^*_{f,i}
        \Big]  \Bigg)
\nn \\ &&
        + \Big( \frac{\xi_{ijk}}{2} \Phi_{s,i} \psi^\top_{s,j} \,e\, 
          \psi_{s,k} 
        + \frac{\xi^*_{ijk}}{2} \Phi^*_{s,i} \psi^\dag_{s,j} \,e\, 
          \psi^*_{s,k} \Big) \, ,
\label{wzlagrangian}
\nd
where the dot product $\cdot$ refers to SU($\nc$) indices and $e$ is the
antisymmetric $2\times 2$ matrix, $e=i\sigma^2$.
Note that the 3-dimensional gamma matrices are $2\times 2$ matrices, and
are in fact nothing but the Pauli matrices, $\sigma^\mu$.

The lattice implementation of this theory is obtained by discretizing
space onto a cubic lattice of spacing $a$, replacing 
$\int d^3 x {\cal L}$ with $a^3 \sum_x {\cal L}$, and writing the
gradient terms as follows:
\bg
\int d^3 x \; (\partial_\mu \Phi)^* (\partial^\mu \Phi)
        & \rightarrow &
        a^3 \sum_{x,\mu} \left( 
        \frac{\Phi^*(x{+}a \hat\mu) - \Phi^*(x)}{a} \;
        \frac{\Phi(x{+}a \hat\mu) - \Phi(x)}{a} \right) \, , \\
\int d^3 x \; \psi^\dag \nott{\partial} \psi
        & \rightarrow &
        a^3 \sum_{x,\mu} \psi^\dag  \, \left[\sigma^\mu 
        \frac{\psi(x{+}a \hat\mu)-\psi(x{-}a \hat\mu)}{2a}
        \right. \nn \\ && \hspace{0.7in} \left. 
        + \frac{a r}{2} \; \frac{-\psi(x{+}a \hat\mu) + 2 \psi(x)
        -\psi(x{-}a \hat\mu)}{a^2} \right] \, .
\nd
The ``extra'' term in the fermion gradient term, which for slowly
varying fields looks like $-(ra/2) \psi^\dag \partial^2 \psi$, is called
the Wilson term, and is required to remove fermion doublers.  It
vanishes in the continuum $a \rightarrow 0$ limit, but its presence
complicates the treatment, particularly because the value of the Wilson
coefficient $r$ is to be chosen by the practitioner (though $r>0$ is
required and $r \leq 1$ is desirable to ensure reflection positivity).  We
will present results for the two values of $r=1$ and $r=\frac 12$.

In addition to these, it is necessary to add the following mass
(counter)terms to the lattice Lagrangian:
\bg
\delta {\cal L} & = & \delta m^2_{f,ij} \Phi^*_{f,i} \Phi_{f,j}
        + \delta m^2_{a,ij} \Phi^*_{a,i} \Phi_{a,j}
        + \delta m^2_{s,ij} \Phi^*_{s,i} \Phi_{s,j}
\nn \\ &&
        + \delta M_{f,ij} \psi^\dag_{f,i} \psi_{f,j}
        + \delta M_{a,ij} \psi^\dag_{a,i} \psi_{a,j}
        + \delta M_{s,ij} \psi^\dag_{s,i} \psi_{s,j} \, .
\nd
In principle we should also allow for multiplicative renormalization of
the couplings and wave functions by including $Z$ factors for each, but
this is not needed at the order of interest.

To write the propagators of the lattice fields it is convenient to
define the quantities
\st
\tilde{p}^2 \equiv \sum_{\mu} \frac{4}{a^2} \sin^2 \frac{p_\mu a}{2}
        \, , \qquad
\hat p_\mu \equiv \frac{1}{a} \sin p_\mu a \, ,
\stp
in terms of which the propagators of the scalar and fermionic fields are 
\st
\Delta_p = \frac{1}{\tilde{p}^2} \, , \qquad
S_p = \frac{-i\sum_\mu \sigma^\mu \hat p_\mu + M_p }
        {\hat p^2 + M_p^2} \, .
\label{propagators}
\stp
Here $M_p= \frac{ar}{2} \tilde{p}^2$ is the momentum dependent 
effective lattice mass induced by the Wilson term.

\FIGURE[t]{
\centerline{\epsfxsize=4in\epsfbox{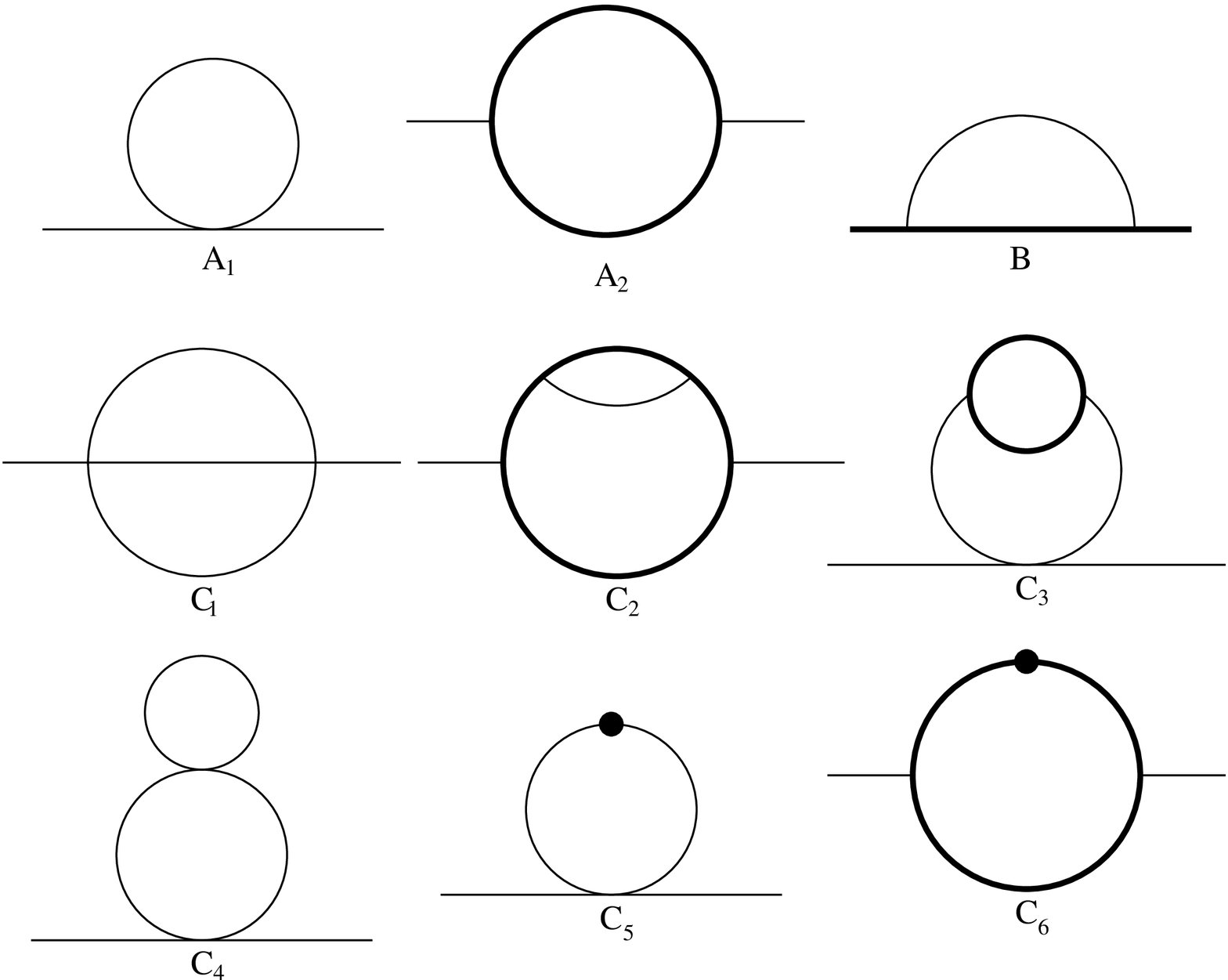}}
\caption{All diagrams needed in scalar and fermionic mass corrections;
$A$ are for $C_{ys}$, $B$ is for $C_{yf}$, and $C$ are for $C_{yy}$.
Thin lines are scalars, heavy lines are fermions.
All of the two loop diagrams have the same combinatorial factor, so the
integrands can be added directly to produce an IR finite overall
integrand.  The heavy dots are one loop mass counterterms, given by the
one loop diagrams evaluated at vanishing external
momentum. \label{fig1}}
}

We need to do three integrals using these lattice propagators to
complete the matching calculation; define
\bg
\frac{C_{ys}}{4\pi a} & \equiv & \int_{\rm BZ} \frac{d^3 p}{(2\pi)^3}
        \Big( 2 \Delta_p + \Tr S_p S_p \Big) \, , \\
\frac{C_{yf}}{4\pi} & \equiv & \int_{\rm BZ} \frac{d^3 p}{(2\pi)^3}
        \, \Delta_p \left(S_p + S_{-p}\right)  \, , \\
\frac{C_{yy}}{16\pi^2} & \equiv & \int_{BZ} \frac{d^3 k d^3
        p}{(2\pi)^6} \Bigg( 
        \Delta_k\Delta_p\Delta_{p{+}k}
        -\Dk \, \Tr\bigg(S_p^3 (S_{k+p}-S_p)
                                +S_{k+p}^3 (S_p-S_{k+p})\bigg) 
                \nn\\ && \hspace{.75in}
        +\Dk^2 \, \Tr\bigg(S_p S_{k+p}
        -\hf(S_p^2+S_{k+p}^2)\bigg)\Bigg)  \, .
\nd
Here $BZ$ means that the integral is to be taken over a cubic Brillouin
zone of extent $(-\pi/a,\pi/a]^3$, that is, each momentum component runs
from $-\pi/a$ to $\pi/a$.  We have included the factors of $4\pi$ to
imitate the expected behavior that each loop order involves $y^2/4\pi$,
and to match notation with previous literature \cite{FKRS2}.
The diagrams responsible for the contributions are presented in Figure
\ref{fig1}; the integrations for $C_{yy}$, the two loop scalar mass
correction, have been written in the way presented so that the integrand
is absolutely integrable and carries all of its singularities on the
three surfaces $k=0$, $p=0$, and $p{+}k=0$, which is
convenient for numerical evaluation.  

We can readily verify that, at low momenta, each integrand is well
behaved.  At leading order for small $p$, the integrand for $C_{ys}$
behaves as
\st
2\Delta(p) + \Tr S^2(p) \simeq \frac{2}{p^2} 
        + \frac{\Tr (i\nott{p})^2}{p^4}
        = \frac{2}{p^2} - \frac{2}{p^2} = 0 \quad
        (\mbox{each term is $+O(p^0)$}) \, .
\stp
Similarly, the integrand of $C_{yf}$ goes as $ra/p^2$ at small $p$.  The
integrand for $C_{yy}$ requires more patience to analyze, but it also
proves to be well behaved for $p \rightarrow 0$, for $k \rightarrow 0$,
and for $p,k$ simultaneously taken to zero.

None of the integrals can be done
in closed form (to our knowledge), but all are relatively tractable by
quadratures; we find numerically,
\st
\begin{array}{ll}
C_{ys}(r=1)  = 6.4706034146527591308  \quad &
C_{ys}(r=0.5) = 5.057247581039541 \\
C_{yf}(r=1)  = 2.29977456857632 \quad &
C_{yf}(r=0.5) = 2.22804716126902  \\
C_{yy}(r=1)  = 5.425954134(5) \quad &
C_{yy}(r=0.5) = 6.8513618(8) \, . \\ \end{array}
\stp
In terms of these coefficients, the required renormalizations of the 
masses are
\bg
\delta m_{s,ij}^2(\mbox{1 loop}) & = & 
        \left( - \df \lambda^*_{lmi} \lambda_{lmj}
        - \frac 12 \xi^*_{ilm} \xi_{jlm} \right) \frac{C_{ys}}{4\pi a}
        \, , \nn \\
\delta m_{f,ij}^2(\mbox{1 loop}) & = & 
        -\lambda^*_{ilm} \lambda_{jlm} \frac{C_{ys}}{4\pi a}
        \, , \nn \\
\delta m_{a,ij}^2(\mbox{1 loop}) & = & 
        -\lambda^*_{lim} \lambda_{ljm} \frac{C_{ys}}{4\pi a}
        \, ,
\nd
for the scalar masses at one loop,
\bg
\delta M_{s,ij} & = & 
        \left( \df \lambda^*_{lmi} \lambda_{lmj}
        + \frac 12 \xi^*_{ilm} \xi_{jlm} \right) \frac{C_{yf}}{4\pi}
        \, , \nn \\
\delta M_{f,ij} & = & 
        \lambda^*_{ilm} \lambda_{jlm} \frac{C_{yf}}{4\pi}
        \, , \nn \\
\delta M_{a,ij} & = & 
        \lambda^*_{lim} \lambda_{ljm} \frac{C_{yf}}{4\pi}
        \, ,
\nd
for the fermions, and
\bg
\delta m_{s,ij}^2(\mbox{2 loop}) & = &  \bigg\{ 
        +\df \lambda_{nmi}^* \lambda_{nqj} \lambda^*_{lqk}
        \lambda_{lmk} + \df \lambda^*_{mni} \lambda_{qnj}
        \lambda^*_{qlk} \lambda_{mlk}
        \nn \\ && \phantom{\bigg\{} 
        + \df \xi^*_{qni} \xi_{qmj} \lambda^*_{klm} \lambda_{kln}
        \;+ \frac 12 \xi^*_{kmi} \xi_{lmj} \xi^*_{knq} \xi_{lnq}
        \;\;\bigg\} \frac{C_{yy}}{16 \pi^2} \, , \nn \\
\delta m^2_{f,ij}(\mbox{2 loop}) & = & \bigg\{
        \lambda^*_{imn} \lambda_{jqn} \lambda^*_{kql} \lambda_{kml}
        \nn \\ && \phantom{\bigg\{} 
        + \df \lambda^*_{inm} \lambda_{jnq} \lambda^*_{klq} \lambda_{klm}
        + \frac 12 \lambda^*_{ikl} \lambda_{jkm} \xi^*_{nqm} \xi_{nql}
        \bigg\} \frac{C_{yy}}{16 \pi^2} \, , \nn \\
\delta m^2_{a,ij}(\mbox{2 loop}) & = & \bigg\{
        \lambda^*_{min} \lambda_{qjn} \lambda^*_{qkl} \lambda_{mkl}
        \nn \\ && \phantom{\bigg\{} 
        + \df \lambda^*_{nim} \lambda_{njq} \lambda^*_{lkq} \lambda_{lkm}
        + \frac 12 \lambda^*_{kil} \lambda_{kjm} \xi^*_{nqm} \xi_{nql}
        \bigg\} \frac{C_{yy}}{16 \pi^2} \, ,
\nd
for the scalar masses at two loops.
Here $\df$ is the dimension of the fundamental representation, $\df=\nc$
in SU($\nc$) gauge theory.  The full scalar mass counterterm is the sum
of the 1 and 2 loop contributions.

This completes the renormalization of the theory at a level which will
leave only $O(a)$ supersymmetry breaking effects.

\section{$N{=}2$ SU($\nc$) gauge theories with fundamental matter}
\label{sec:gauge}

Now we extend these results to the case where gauge interactions are
also present.  We will not review here, how gauge fields are put on the
lattice; we refer the interested reader to H.\ Rothe's book
\cite{Rothe}, which also presents the Feynman rules associated with the
pure gauge and gauge/fermion sectors with 
$1/4 \rightarrow C_A/12$ in Eq.\ (15.39) and 
$2/3(\delta_{AB}\delta_{CD} + \dots) \rightarrow 
\sum tr(T^A T^B T^C T^D)$ in Eq.\ (15.53b)
for the generalization to SU($\nc$) (the sum is over all 
permutations of ABCD).

The added fields are a gauge field
$A_\mu$, an adjoint fermionic gaugino $\chi$, and an adjoint scalar
field we will write $\phi$.  In the approach where we obtain a 3-D $N{=}2$
supersymmetric theory by dimensional reduction of a 4-D, $N{=}1$ SUSY
theory, the scalar $\phi$ is the gauge field component in the direction
which was compactified.  Besides their
gauge interactions, which follow from each species' gauge
representation, these fields also introduce new Yukawa and scalar
couplings,
\bg
\!\!{\cal L}(\mbox{new}) & = & \sqrt{2}g \Big( 
        \psi^\dag_{f,i} \,e\, (\chi^A)^* \, T^A \, \Phi_{f,i}
        + \Phi^*_{f,i} \,T^A \, \psi^\top_{f,i} \,e\, \chi^A
        + (f \rightarrow a) \Big)
        \nn \\ && 
        +\frac{g^2}{2} \Big( \Phi^*_{f,i} T^A \Phi_{f,i} 
        + \Phi^*_{a,i} T^A \Phi_{a,i} \Big)
        \Big( \Phi^*_{f,j} T^A \Phi_{f,j} 
        + \Phi^*_{a,j} T^A \Phi_{a,j} \Big)
        \nn \\ &&
        + g^2 \phi^A \phi^B \Big(
        \Phi^*_{f,i} T^A T^B \Phi_{f,i} 
        + \Phi^*_{a,i} T^A T^B \Phi_{a,i} \Big) 
        \nn \\ &&
        +g\phi^A \Big( \psi^\dag_{f,i} T^A \psi_{f,i} + 
        \psi^\dag_{a,i} T^A \psi_{a,i} \Big)
        +g\phi^A \chi^\dag F^A \chi \, ,
\label{extralagrangian}
\nd
where $F^A_{BC}=-if_{ABC}$ is the generator of SU($\nc$) in the adjoint 
representation and $T^A$ must be treated in 
context as the group generator 
in the fundamental representation when between $f$ species, or the
antifundamental representation ($-T^*$) when between $a$ species.  These
interactions, which are local, can be implemented on the lattice in the
obvious way.  The gauge interaction implementation is not so simple but
has been well treated in the literature. 

Though the matching calculation as a whole is perfectly IR safe and 
gauge invariant, individual diagrams are, in general, not. The 
calculation reduces to determining IR safe combinations of diagrams that 
are gauge invariant. To this end we define the invariants of the group 
representation as 
\bg
\Tr \left(T^AT^B\right)
        & \equiv & {\cal T}_F\delta^{AB}
\nn\\
\left(T^AT^A\right)_{ab} 
        & \equiv & {\cal C}_F\delta_{ab}
\nn\\
\Tr\left(F^AF^B\right) & = & \left(F^CF^C\right)_{AB}
        \equiv{\cal C}_A\delta_{AB}\, .
\nd
Here ${\cal T}_F {\equiv} C(\nc)$ is the first invariant of the 
fundamental (defining) representation 
of SU($\nc$) also called the trace normalization. It is usually 
chosen to be $\hf$. With that choice the second invariant is  
${\cal C}_F {\equiv} C_2(\nc){=}(\nc^2{-}1)/2\nc$. ${\cal C}_A{\equiv} C(G)
{=}C_2(G){=}\nc$ 
are the invariants of the adjoint 
representation of the group. $n_f$ and $n_a$ are the number of 
fundamental and anti-fundamental multiplets respectively.
Each diagram will include a prefactor involving some combination of these 
invariants, the Yukawa couplings, the dimension of the fundamental 
representation and the sum $n_f+n_a$. Since the matching calculation is 
valid for {\em any} Lie group, we can factorize the diagrams based on 
this prefactor.

Two other group theoretic relations are needed to complete the calculation, 
which we provide here since they appear in the results almost unchanged:
\bg
T^BT^AT^B = \left({\cal C}_F-\hf{\cal C}_A\right)T^A 
\qquad \mbox{and} \qquad F^A_{BC}T^BT^C=\hf{\cal C}_AT^A \, .
\nd

%

New counterterms are needed, which we name $\hf\delta m^2_\phi$ and 
$\delta M_\chi$; they have the obvious definitions.
The new contributions to the one loop mass counterterms turn out to
require only one new lattice integral,
\bg
\frac{C_{gf}}{4\pi} \equiv \int_{BZ}\frac{d^3 p}{(2\pi)^3}
                \Bigg( -\frac{r}{2}\Delta_p + \frac{1-r^2}{4} 
                \left(\frac{M_p}{\hat{p}^2+M_p^2}\right) \Bigg)
\nd
which is known analytically for $r=1$ and easily determined for $r=\hf$:
\st
\begin{array}{ll}
C_{gf}(r=1)  = - \Sigma /2 \quad &
C_{gf}(r=0.5) = .097938749331668  \\ \end{array}
\stp
where $\Sigma=3.17591153562522$ \cite{FKRS2}.  They are,
\bg
        \left(\delta m^2_{a,f}\right)_{ab} & = & 
        -2g^2 {\cal C}_F\,\delta_{ab} \frac{C_{ys}}{4\pi a}
                \nn \, , \\
        \left(\delta m^2_{\phi}\right)_{AB} & = & 
        - g^2\bigg((n_f+n_a){\cal T}_F+{\cal C}_A\bigg)
        \delta_{AB}\frac{C_{ys}}{4\pi a}
                \nn \, , \\
        \left(\delta M_{a,f}\right)_{ab} & = & 
        g^2 {\cal C}_F\,\delta_{ab} \frac{C_{gf}}{4\pi}
                \nn \, , \\
        \left(\delta M_{\chi}\right)_{AB} & = & g^2  
        {\cal C}_A \frac{C_{gf}}{4\pi} +g^2
         \bigg((n_f+n_a){\cal T}_F+{\cal C}_A\bigg)\delta_{AB}\frac{C_{yf}}{4\pi}
                 \, . 
\nd

The new two loop counterterms come in many combinations factorized 
according to the prefactors determined by the group theory (i.e. 
products of the group invariants).
In terms of the nine new integrals defined in appendix \ref{sec:int} 
the new counterterms are (obviously 
$\lambda^*_{ilm}\lambda_{jlm}\to\lambda^*_{lim}\lambda_{ljm}$ 
for $\delta m^2_a$ in the second expression)
 \bg
     \delta m^2_{s,ij} &=&
         g^2 \lambda^*_{lmi}\lambda_{lmj}{\cal C}_F d_F 
                \frac{C^{sing}_{g}}{16\pi^2} 
                          \,\, , \\
     \left( \delta m^2_{a,f;ij} \right)_{ab}&=&
         g^2\delta_{ab}\lambda^*_{ilm}\lambda_{jlm}
        {\cal C}_F\frac{C^{fund}_{g1}}{16\pi^2}         
       +\,  g^4 \delta_{ij} \delta_{ab}\Bigg\{
           {\cal T}_F {\cal C}_F (n_f + n_a)\frac{C^{fund}_{g2}}{16\pi^2}
                        \nn\\ &&
        +\big( {\cal C}_F\big)^2 \left(\frac{C^{fund}_{g3}}{16\pi^2}
                      - \frac{1}{3}\frac{\Sigma^2}{16\pi^2} \right)
         + {\cal C}_F {\cal C}_A \left( \frac{C^{fund}_{g4}}{16\pi^2} 
                      +\frac{1}{18} \frac{\Sigma^2}{16\pi^2} \right)
                        \nn\\ &&
         - \frac{4}{3} {\cal T}_F {\cal C}_F 
                \bigg( {\cal C}_F - \frac{1}{6} {\cal C}_A\bigg) 
                \frac{(4\pi\Sigma)}{16\pi^2} \Bigg\}     
                                \,\, ,\qquad  \\
     \left( \delta m^2_{\phi} \right)_{AB} &=&
         g^2\delta_{AB} \lambda^*_{ijk}\lambda_{ijk} 
        {\cal T}_F\frac{C^{adj}_{g1}}{16\pi^2}
        + \, g^4 \delta_{AB} \Bigg\{
           {\cal T}_F {\cal C}_F (n_f + n_a)\frac{C^{adj}_{g2}}{16\pi^2}
         + {\cal T}_F {\cal C}_A \frac{C^{adj}_{g3}}{16\pi^2} 
                                \nn \\ && 
       + \big({\cal C}_A\big)^2 \left( \frac{C^{adj}_{g4}}{16\pi^2}
                 - \frac{5}{18} \frac{\Sigma^2}{16\pi^2} \right)
        - \frac{4}{3} {\cal T}_F {\cal C}_A \bigg( 
        {\cal C}_F - \frac{1}{6} {\cal C}_A \bigg) 
        \frac{(4\pi\Sigma)}{16\pi^2}\Bigg\} \,\, . \qquad
\label{eqn:result}
 \nd
The last term in each scalar mass correction is from 
the piece of the 4-point gluon vertex with a group structure that is
fully symmetric between the four lines.
It is separately gauge invariant and IR finite, and (as shown) can be
found in closed form in terms of the constant $\Sigma$, 
defined in Eq.\ \ref{eqn:sigma}. 
The group theoretic factor preceding it arises as
$\delta_{AB} \delta_{CD} \sum_{\mbox{\tiny perm}} T^A T^B T^C T^D$.

The constants appearing in these expressions are
\st
\begin{array}{ll}
C^{sing}_{g}(r=1)  = 3.588328893(6)\qquad &
C^{sing}_{g}(r=0.5)  =  17.8901895(7)\\
C^{fund}_{g1}(r=1) = -4.89236097(1)\qquad &
C^{fund}_{g1}(r=0.5) = 3.648535(2)\\
C^{fund}_{g2}(r=1) =  10.2296763(2)\qquad &
C^{fund}_{g2}(r=0.5) = 13.32776(1)      \\
C^{fund}_{g3}(r=1) =  22.712647140(8)\qquad &
C^{fund}_{g3}(r=0.5) =  29.816565(2)\\
C^{fund}_{g4}(r=1) =  -2.647013(1) \qquad &
C^{fund}_{g4}(r=0.5) =  9.051300(7)     \\
C^{adj}_{g1}(r=1)  =  7.75588650(2)\qquad &
C^{adj}_{g1}(r=0.5)  =  14.526578(3)    \\
C^{adj}_{g2}(r=1)  =  17.536258926(7) \qquad &
C^{adj}_{g2}(r=0.5)  =  30.769058(1)    \\
C^{adj}_{g3}(r=1)  =  -.3347923(2)\qquad &
C^{adj}_{g3}(r=0.5)  = .76791(1)        \\
C^{adj}_{g4}(r=1)  =  13.0938429(1)\qquad &
C^{adj}_{g4}(r=0.5)  =  20.658655(8) \, .\end{array}
\stp
The number in parentheses is the error in the last digit. It was 
determined conservatively since the accuracy of these constants is 
not expected to be a limiting factor in any lattice implementation.
This completes the renormalization of the theory at a level which will
leave only $O(a)$ supersymmetry breaking effects.



\section{Conclusion}
\label{sec:conclusion}

Non-perturbative treatments of supersymmetric theories have been an 
elusive goal for decades. A recent surge in both interest and progress 
in this field has created a great amount of excitement in both the 
theoretical and lattice physics communities. However, these theories 
are only recently showing results in terms of the study of truly 
non-perturbative SUSY phenomena. Much of this difficulty has to do 
with the great technical challenges encountered in the attempt to 
formulate lattice theories that can reproduce SUSY in the IR. The 
direction of the field up until this point has been to try to build
infrared supersymmetric behavior into the construction of the action.
We have argued that, for three dimensional theories, it is feasible and
straightforward instead to use the simplest possible action and do the
fine tuning of its parameters necessary to obtain infrared
supersymmetric behavior.  This is possible analytically and does not
prove as difficult as one might have feared.

We have performed these tunings 
for a class of theories displaying $N{=}2$ 
supersymmetry in three dimensions and containing arbitrary numbers of 
matter multiplets transforming in the fundamental representation of the
gauge group.
The technique is robust in the sense that it relies only on theoretical 
principles, like Wilson's effective action formulation, and lattice 
implementations, like the Wilson action for fermions, that have 
been rigorously studied for decades. More generally, the entire 
lattice action is
the most simple such construction with the appropriate IR limit, 
such that the many complications that can arise in such theories are 
suitably manageable or altogether absent. 

Though we have not done so, it should be very straightforward to extend
our results to matter in general representations.  This allows, for
instance, the extension of our results to SYM theories with
higher supersymmetry, in which interesting
non-perturbative physics is claimed to exist.  For instance,
the $N{=}8$ SYM theory in three dimensions has been conjectured 
by Seiberg to possess a non-trivial IR fixed point (an interacting 
conformal theory) \cite{Seiberg:1997ax}. The theory can be constructed 
as the $N{=}2$ SYM theory with 3 complex matter hyper-multiplets of 
the $N{=}2$ theory 
all transforming under the adjoint representation of the gauge group. 
A lattice has recently been constructed in \cite{Kaplan:2005ta} to 
study this theory by a very different technique. 

\begin{acknowledgments}

We would like to acknowledge discussions with David Kaplan (which
initiated this work), and with Matt Strassler.
This work was supported, in part, by 
the National Sciences and Engineering
Research Council of Canada, and by le Fonds Nature et Technologies du
Qu\'ebec.

\end{acknowledgments}

\appendix

\FIGURE[ht]{
\centerline{\epsfxsize=5in\epsfbox{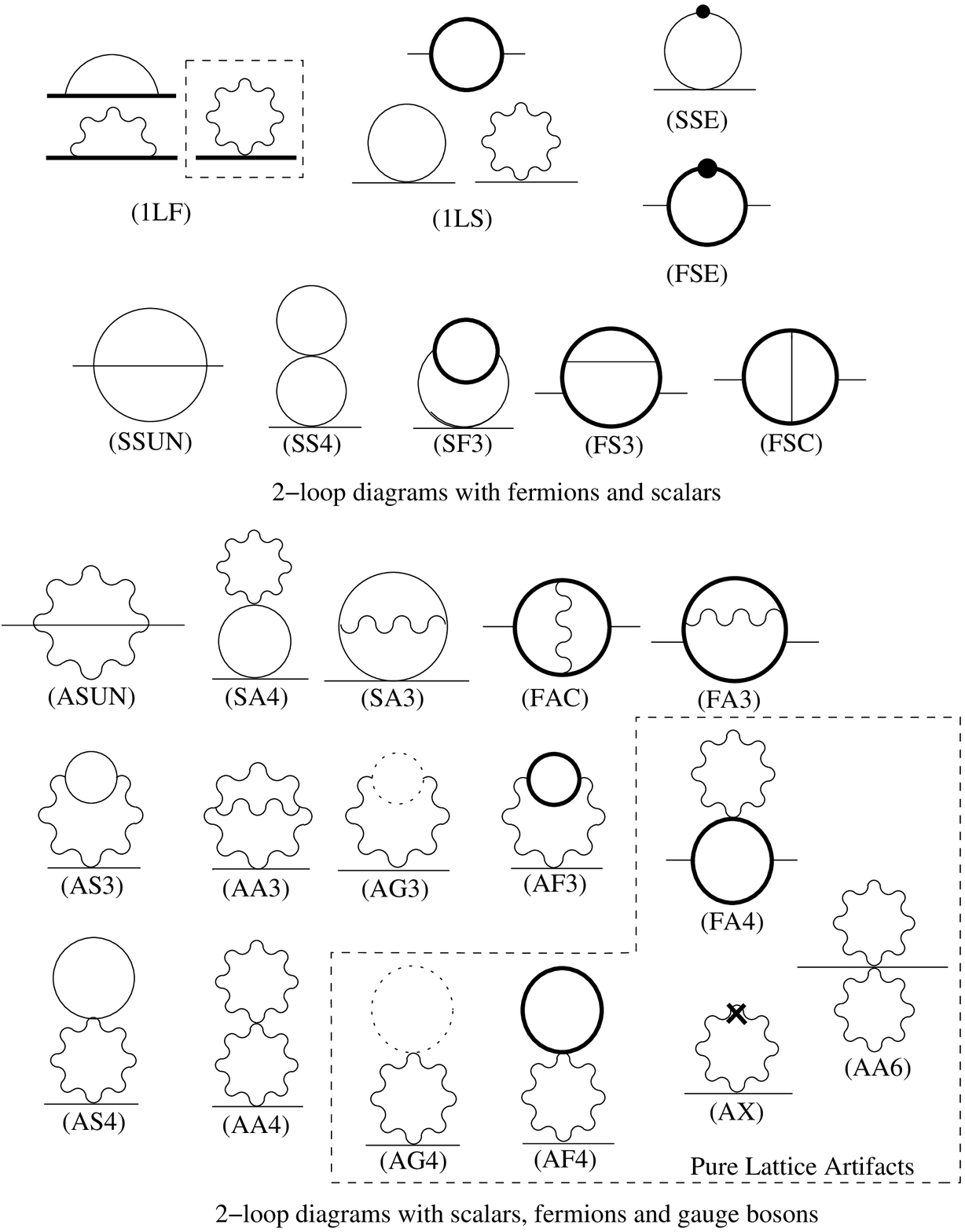}}
\caption{Diagrams needed for the renormalization of SQCD.
\label{fig2}}
}

\section{Diagrams and Integrals}
\label{sec:int}

In this appendix we enumerate precisely the integrals involved in the 
two-loop mass correction. All necessary diagrams are given in Fig. 
\ref{fig2}. Heavy lines are fermions, light lines are scalars, 
dotted lines are ghosts and wavy lines are gauge bosons. Heavy dots 
represent the one-loop mass insertion counter-term, the one-loop diagrams 
evaluated at zero external momentum. The dark cross in diagram (AX) 
is the measure counter-term. 
Diagrams that arise purely as artifacts of the lattice implementation 
are displayed in a dashed box. Diagrams that do not contribute in Landau 
gauge are not displayed.

Since it makes structural sense to do so, 
we have grouped the expression for each two-loop diagram with the
analogous mass insertion counter-term. The abbreviations on the LHS 
correspond to the labels from Fig. \ref{fig2}. The notation is 
relatively straightforward: the first letter (F, S or A) labels the 
field which couple directly to the external scalar line (fermion, 
scalar or gauge boson respectively), the 
second, except in the case of the sunset diagrams and (AA6), 
labels the field coupling secondarily and the number shows whether 
that secondary coupling is via a 3-point or 4-point vertex (replaced by 
C when the secondary line ``crosses'' the diagram). X represents the 
measure counter-term, a pure artifact of the lattice construction that 
comes about because the link variable integration measure in the 
path integral has non-trivial dependence on the gauge fields \cite{Rothe}.
G is the Fadeev-Popov ghost, which arises from gauge fixing in direct 
analogy to the continuum except that the ghost-gauge interaction 
contains an infinite number of new irrelevant vertices. Of these, 
only the GGAA vertex will contribute to continuum correlation functions 
at the perturbative order we require. Its contribution looks like 
a non-renormalizable mass divergence which is essential to 
ensure that these divergences cancel in the correlation 
functions, as is guaranteed by gauge invariance.

The integrands involving only scalars and fermions are (not 
including $(SS4)$ since its contribution is exactly canceled 
by its analogous mass counter-term insertion)
\bg
(SSUN) &=& \Dk \Dp \Dq \nn\\
(SF3)-c.t. &=& \Dk^2 \, \Tr\bigg(S_p S_{k{+}p}
        -\hf(S_p^2+S_{k{+}p}^2)\bigg) \nn\\
(FS3)-c.t. &=& \hf \Dk \, \Tr\bigg(S_p^3 (S_{k{+}p}-S_p)
                        +S_{k{+}p}^3 (S_p-S_{k{+}p})\bigg) \nn\\
(FSC) &=& \Dk \, \Tr(S_p^2 S_{k{+}p}^2) \, .
\nd
For the scalar-gauge and ghost-gauge sectors we need to define the 
gauge boson propagator on the lattice (Landau gauge is used throughout 
the calculations):
\bg
\tilde{\Delta}^{\mu\nu}_k=\Delta_k\left(\delta_{\mu\nu}- 
  \frac{\tilde{k}_\mu\tilde{k}_\nu}{\tilde{k}^2}\right)\, ,
\nn
\nd
in terms of which they are
\bg
(SA3) &=& \hf \tilde{\Delta}^{\mu\nu}_k \Dp\Dq (\Dp+\Dq)
        \widetilde{(2p{+}k)}_\mu \widetilde{(2p{+}k)}_\nu \nn\\
(SA4)-c.t. &=& \hf \tilde{\Delta}^{\mu\nu}_k
        \bigg(\Dp^2(\cos(p_\mu)-1) +  \Dq^2(\cos(k{+}p)_\mu-1)\bigg)
        \delta_{\mu\nu} \nn\\
(ASUN) &=& \Dk \tilde{\Delta}^{\mu\nu}_p 
        \tilde{\Delta}^{\alpha\beta}_{k{+}p}
        \cos(\hfk)_\mu \cos(\hfk)_\nu \,
        \delta_{\mu\alpha}\delta_{\nu\beta}\nn\\
(AS3) &=& \tilde{\Delta}^{\mu\nu}_k \tilde{\Delta}^{\alpha\beta}_k
        \Dp\Dq \widetilde{(2p{+}k)}_\mu \widetilde{(2p{+}k)}_\alpha
        \delta_{\nu\beta} \nn\\
(AS4) &=& \hf \tilde{\Delta}^{\mu\nu}_k \tilde{\Delta}^{\alpha\beta}_k
        \bigg(\Dp \cos(p_\mu)+\Dq \cos(k{+}p)_\mu\bigg)
        \delta_{\mu\alpha}\delta_{\nu\beta}\nn\\
(AG3) &=& \tilde{\Delta}^{\mu\nu}_k \tilde{\Delta}^{\alpha\beta}_k
        \Dp\Dq \widetilde{(k{+}p)}_\mu\cos(\hfp)_\mu 
        \,\tilde{p}_\alpha \cos(\hfq)_\alpha \,
        \delta_{\nu\beta} \nn\\
(AG4) &=& \frac{1}{24}\tilde{\Delta}^{\mu\nu}_k 
        \tilde{\Delta}^{\alpha\beta}_k
        \bigg(\Dp^2 \, \tilde{p}_\mu \tilde{p}_\mu 
        + \Dq^2 \widetilde{(k{+}p)}_\mu\widetilde{(k{+}p)}_\mu\bigg)
        \delta_{\mu\alpha}\delta_{\nu\beta} \, .
\label{scalar/gauge}
\nd
For the fermion-gauge sector it is convenient to define the momentum 
part of the $\psi\psi A$ and $\psi\psi AA$ vertices as
\st
V^\mu_p = \sigma_\mu \cos(\hfp)_\mu - \frac{ir}{2}\tilde{p}_\mu
                \qquad \mbox{and}\qquad 
V^{'\mu}_p = -\frac{i}{2}\sigma_\mu \tilde{p}_\mu +r\cos(\hfp)_\mu
\stp
respectively. In terms of these the integrands take the form 
\bg
(FAC) &=& \tilde{\Delta}^{\mu\nu}_k \, \Tr\bigg(S^2_p V^\mu_{2p{+}k}
                                S^2_{k{+}p}V^\nu_{2p{+}k}\bigg)\nn\\
(FA3)-c.t. &=& \hf\tilde{\Delta}^{\mu\nu}_k 
        \, \Tr\bigg(S^3_p \Big(V^\mu_{2p{+}k}S_{k{+}p}V^\nu_{2p{+}k}
                        -\hf[V^\mu_kS_k V^\nu_k
                        + V^\mu_{-k}S_{-k} V^\nu_{-k}] \Big)\nn\\
                &&\qquad\qquad
                +S^3_{k{+}p} \Big(V^\mu_{2p{+}k}S_p V^\nu_{2p{+}k}
                        -\hf[V^\mu_kS_k V^\nu_k
                        + V^\mu_{-k}S_{-k} V^\nu_{-k}]\Big)\bigg)\nn\\
(FA4)-c.t. &=& \hf\tilde{\Delta}^{\mu\nu}_k 
        \Tr\bigg(S^3_p (V^{'\mu}_{2p}-r)
                +S^3_{k{+}p}(V^{'\mu}_{2k{+}2p}-r)\bigg)
        \delta_{\mu\nu} \nn\\
(AF3) &=& \tilde{\Delta}^{\mu\nu}_k \tilde{\Delta}^{\alpha\beta}_k
        \Tr S_p  V^\mu_{2p{+}k} S_{k{+}p} V^\alpha_{2p{+}k}
        \,\delta_{\nu\beta} \nn\\
(AF4) &=& \hf\tilde{\Delta}^{\mu\nu}_k \tilde{\Delta}^{\alpha\beta}_k
        \Tr \bigg(S_p V^{'\mu}_{2p}
                +S_{k{+}p}V^{'\mu}_{2k{+}2p}\bigg)
        \delta_{\mu\alpha}\delta_{\nu\beta} \, .
\nd

Finally, for the pure gauge sector, we have
\bg
(AX)  &=&  \frac{1}{12}\tilde{\Delta}^{\mu\nu}_k 
                \tilde{\Delta}^{\alpha\beta}_k 
                \delta_{\mu\alpha}\delta_{\nu\beta}
                 =\frac{1}{6}\Delta^2_k
         \nn\\
(AA3) &=& \tilde{\Delta}^{\mu\eps}_k \tilde{\Delta}^{\alpha\eps}_k
        \tilde{\Delta}^{\nu\beta}_p 
        \tilde{\Delta}^{\lambda\gamma}_{k{+}p}
        \nn\\ &&\!\!\!\!\!\!\!\!\!\!\!\!\!\!\times
   \bigg\{{-}\delta_{\nu\lambda}\widetilde{(2p{+}k)}_\mu\cos(\hfk)_\nu
        +\delta_{\mu\lambda}\widetilde{(2k{+}p)}_\nu\cos(\hfp)_\lambda
        +\delta_{\mu\nu}\widetilde{(p{-}k)}_\lambda\cos(\hfq)_\mu
        \bigg\}\nn\\ &&\!\!\!\!\!\!\!\!\!\!\!\!\!\!\times
\bigg\{{+}\delta_{\beta\gamma}\widetilde{(2p{+}k)}_\alpha\cos(\hfk)_\beta 
-\delta_{\alpha\gamma}\widetilde{(2k{+}p)}_\beta\cos(\hfp)_\gamma
 -\delta_{\alpha\beta}\widetilde{(p{-}k)}_\gamma\cos(\hfq)_\alpha
        \bigg\} \, . \qquad
\nd

The diagram labelled (AA6) is an artifact of the lattice discretization. 
Its contribution can be solved for exactly in terms of the complete 
elliptic integral of the first kind. Borrowing some now standard 
notation from \cite{FKRS2} we define 
\st
\Sigma \equiv \frac{1}{\pi^2}\int^{\pi/2}_{\pi/2}d^3x
        \frac{1}{\sum_i\sin^2(x_i)}=3.17591153562522 \, .
\label{eqn:sigma}
\stp
It is
\bg
(AA6)= - \frac{1}{4} \bigg[\frac{2}{3}t^At^At^Bt^B
                        +\frac{1}{3}t^At^Bt^At^B\bigg]
        \int_{k,p} \tilde{\Delta}^{\mu\nu}_k\tilde{\Delta}^{\mu\nu}_p\, .
        \nn
\nd
Repeated Lorentz indices are summed. $t^A$ represents the generator 
of SU($\nc$) in either the defining ($T^A_{ab}$) 
or adjoint ($F^A_{BC}$) 
representation depending on the transformation properties of the 
external scalar in question. Appropriate external indices for the 
product of $t$'s is implied. The group factor in square brackets 
reduces with the standard relations to
\bg
{\cal C}_F\left({\cal C}_F-\frac{1}{6}{\cal C}_A\right)
&\qquad& \mbox{for external}\,\, \Phi_{a,f}\nn\\\mbox{or}\qquad\qquad
\frac{5}{6}\left({\cal C}_A\right)^2 
&\qquad& \mbox{for external}\,\, \phi\nn \,\,\, .
\nd
The momentum integral is 
\st
\int_{k,p}\Delta_k\Delta_p\left(1+
        \frac{(\tilde k\cdot\tilde p)^2}{\tilde{k}^2\tilde{p}^2}\right)
\, ,
\stp
where $\tilde{k}\cdot\tilde{p}= 4\sum_i\sin(k_i/2)\sin(p_i/2)$. The 
cross terms from the square are odd in the integration variables and 
thus integrate to zero in the Brillouin zone. The numerator of the
second term can then be rewritten as an average, since the integral is 
unchanged by $p_1\leftrightarrow p_2\leftrightarrow p_3$, and 
we see that the exact answer
\bg
\int_{k,p}\tilde{\Delta}^{\mu\nu}_k\tilde{\Delta}^{\mu\nu}_p
        = \frac{4}{3}\int_k\Delta_k\int_p\Delta_p
        =\frac{4}{3}\left(\frac{\Sigma}{4\pi}\right)^2\nn\,.
\nd

Last, though certainly not least, we have the 
diagram (AA4). We will not reproduce the 4-point gluon vertex rule here 
but it is written out in \cite{Rothe} (with the changes mentioned 
in Sec. \ref{sec:gauge}). The diagram contains two pieces.  One involves
the group structure $f_{ABE} f_{CDE}$, as in the continuum, and
contributes to $C^{fund}_{g4}$ 
and $C^{adj}_{g4}$.  There is also a pure lattice artifact piece with
group structure ${\rm Tr}\: T^A T^B T^C T^D$ plus permutations.  The
contribution of this piece is separately gauge invariant and infrared
finite; it can be computed in a way similar to diagram (AA6) above.

In terms of these we need to do 9 numerical integrals. These are 
composed into IR safe gauge invariant combinations as described in 
Sec. \ref{sec:gauge}. The numerical factors on the diagrams are a 
combination of many different factors. Some care 
must be taken in determining 
extra factors of $(-1)$ from diagrams where $e$ commutes across 
odd numbers of gamma matrices and where odd numbers of the conjugate 
generator $-T^*$ appear. This occurs most commonly in diagrams with 
crossed lines. A notable exception is the ${-}2(SSUN)$ contribution 
in the first line of Eq.\ \ref{eqn:sum2} but this contribution, 
when lines representing 
the auxiliary fields are included, also takes the form of a crossed 
diagram. 

For the correction to the singlet scalar the appropriate IR safe 
gauge invariant combination is (we have changed 
notation so that analogous mass insertion counterterms are 
included implicitly by the diagram labels)
\st
       \frac{C^{sing}_{g}}{16\pi^2} \equiv 
                        4(SF3)-6(FS3)+(FSC)-2(SA3)
                        + (FAC) + 2(FA3)+(FA4)
                          \, .
\label{eqn:sum1}
\stp
For the correction to the fundamental/anti-fundamental scalars 
they are
\bg
         \hspace{-.1in}
        \frac{C^{fund}_{g1}}{16\pi^2} &\equiv&  
                        3(SF3)-7(FS3)+4(FSC)-2(SSUN)
                        -(SA3) + (FA3)+\hf(FA4)
                         \, , \nn\\
         \frac{C^{fund}_{g2}}{16\pi^2} &\equiv& 
                        (SF3)-4(FS3)+(SSUN)-(AF3)
                        -(AF4)-(AS3)+2(AS4)
                          \, , \nn\\
         \frac{C^{fund}_{g3}}{16\pi^2} &\equiv& 
                        2(SF3)-6(FS3)+3(SSUN)-(SA3)
                        + 2(ASUN)+2(FA3)+(FA4)
                          \, , \nn\\
         \frac{C^{fund}_{g4}}{16\pi^2} &\equiv& 
                        (SF3)-2(FS3)+(FSC)-(SSUN)-(SA3)+(FAC)
                        \nn\\&& 
                        -\hf(ASUN)+2(FA3)
                        +(FA4)-\hf(AS3)+(AS4)-(AF3)
                        \nn\\&&
                        -(AF4) +(AG3)-2(AG4)+\hf(AA3)+\hf(AA4)+2(AX)
                         \, ,
\label{eqn:sum2}
\nd
and for the adjoint scalar, 
 \bg
         \frac{C^{adj}_{g1}}{16\pi^2} &\equiv& 
                        4(SF3)-8(FS3)+2(FSC)
                          \, ,\nn\\
         \frac{C^{adj}_{g2}}{16\pi^2} &\equiv& 
                        4(SF3)-6(FS3)-(FSC)+4(SSUN)
                        -2(SA3)
                        \nn\\&&
                        +(FAC)+2(FA3)+(FA4)
                          \, ,\nn\\
         \frac{C^{adj}_{g3}}{16\pi^2} &\equiv& 
                        {-}4(FS3)+\frac{5}{2}(FSC)-(SSUN)-\hf(FAC)
                        \nn\\&&
                        -(AF3)-(AF4)-(AS3)+2(AS4)
                         \, ,\nn\\
         \frac{C^{adj}_{g4}}{16\pi^2} &\equiv& 
                        {-}2(FS3)-\hf(FSC)
                        +\frac{3}{2}(ASUN)+\hf(FAC)+2(FA3)+(FA4)
                        \nn\\&&
                        -\hf(AS3)+(AS4)-(AF3)-(AF4)+(AG3)-2(AG4)
                        \nn\\&&
                        +\hf(AA3)+\hf(AA4)+2(AX)
                        \, .\nn\\
\label{eqn:sum3}
 \nd
These combinations are observed to be perfectly IR safe through 
numerical analysis. Note that, as mentioned previously, the label 
($AA4$) contains only that part of the 4-point gluon vertex from 
\cite{Rothe} that reproduces the appropriate expression in the 
continuum. The other symmetric piece, described in Sec. 
\ref{sec:gauge} just after Eq.\ \ref{eqn:result}, has a unique 
group structure and is handled separately.

\end{document}